\pdfoutput=1
\documentclass[11pt,a4paper]{article}
\usepackage{pos}
\usepackage[utf8]{inputenc}

\usepackage{mathtools}
\usepackage[T1]{fontenc}
\usepackage{bbm}
\DeclareMathAlphabet{\mathbfi}{OML}{cmm}{b}{it}

\let\originalleft\left
\let\originalright\right
\renewcommand{\left}{\mathopen{}\mathclose\bgroup\originalleft}
\renewcommand{\right}{\aftergroup\egroup\originalright}

\makeatletter

\def\@spliteq#1{\begin{equation}\begin{split}#1\end{split}\end{equation}}
\def\splitequation{\collect@body\@spliteq}

\makeatother

\renewcommand{\vec}[1]{{\ifnum9<1#1\mathbf{#1}\else\ifcat\noexpand#1\relax\boldsymbol{#1}\else\mathbfi{#1}\fi\fi}}

\newcommand{\mathi}{\mathrm{i}}
\newcommand{\total}{\mathop{}\!\mathrm{d}}
\newcommand{\1}{\mathbbm{1}}
\newcommand{\eqend}[1]{\,#1}

\makeatletter
\def\expect{\@ifnextchar[{\@expecttw@}{\@expect@ne}}
\def\@expecttw@[#1]#2{\left\langle{#2}\right\rangle^\text{#1}}
\def\@expect@ne#1{\left\langle{#1}\right\rangle}
\makeatother

\allowdisplaybreaks

\title{Trace anomaly of Weyl fermions in the Breitenlohner--Maison scheme for $\gamma_*$}
\ShortTitle{Trace anomaly of Weyl fermions}
\author*[a]{S.~Abdallah}
\author*[b,c]{S.~A.~Franchino-Vi{\~n}as}
\author[a]{M.~B.~Fröb}

\affiliation[a]{Institut f{\"u}r Theoretische Physik, Universit{\"a}t Leipzig,\\Br{\"u}derstra{\ss}e 16, 04103 Leipzig, Germany.}
\affiliation[b]{Departamento de F{\'\i}sica, Facultad de Ciencias Exactas, Universidad Nacional de\\ La Plata, C.C.\ 67 (1900), La Plata, Argentina.}
\affiliation[c]{Institut f{\"u}r Theoretische Physik, Universit{\"a}t Heidelberg,\\Philosophenweg 19, 69120 Heidelberg, Germany.}

\emailAdd{abdallah@itp.uni-leipzig.de}
\emailAdd{s.franchino@thphys.uni-heidelberg.de}
\emailAdd{mfroeb@itp.uni-leipzig.de}

\FullConference{%
  *** The European Physical Society Conference on High Energy Physics (EPS-HEP2021), ***\\
  *** 26-30 July 2021 ***\\
  *** Online conference, jointly organized by Universität Hamburg and the research center DESY ***
}

\abstract{We revisit the conformal anomaly for a Weyl fermion in four dimensions that has generated some debate recently. We employ a perturbative expansion for the metric around Minkowski space, dimensional regularization and a Breitenlohner--Maison prescription for the chiral $\gamma$ matrix.
We obtain a vanishing odd-parity contribution for Weyl fermions in four dimensions, while the even-parity contribution is exactly half the one for a Dirac fermion. }

\begin{document}
\maketitle

\section{Introduction}
\label{sec:introduction}
Since the foundational papers of Adler \cite{Adler:1969gk} and Bell and Jackiw \cite{Bell:1969ts}, it has been known that classical symmetries may be broken at the quantum level because of the renormalization process. In particular, the breaking of conformal symmetry and the resulting trace anomaly of the stress(-energy-momentum) tensor, first shown in \cite{Capper:1974ic}, opens an interesting scenario for astroparticle physics and the inflationary paradigm \cite{Dolgov,Fabris:1998vq,Hawking:2000bb,CesareSilva:2020ihf}. 

Recently, building on some previous ideas regarding second-order differential operators for fields of arbitrary spin \cite{Christensen:1978md} and possible CP violations \cite{Nakayama:2012gu} (see also the new results in \cite{Nakagawa:2021wqh}), it has been suggested that the stress tensor of Weyl fermions in curved space may display a particular trace anomaly \cite{Bonora:2014qla,Bonora:2015nqa,Bonora:2017gzz}. 
By dimensional considerations, the most general form for the anomaly in four dimensions is given by
\begin{equation}
\label{eq:anomaly_general}
g_{\mu\nu} \expect{T^{\mu\nu}} = w C^{\mu\nu\rho\sigma} C_{\mu\nu\rho\sigma} + b \mathcal{E}_4 + c \nabla^2 R + f \epsilon^{\mu\nu\rho\sigma} R_{\mu\nu\alpha\beta} R_{\rho\sigma}{}^{\alpha\beta} \eqend{,}
\end{equation}
where $\expect{T^{\mu\nu}}$ is the vacuum expectation value (VEV) of the stress tensor, $C_{\mu\nu\rho\sigma}$ is the Weyl tensor, $\mathcal{E}_4 = R^2 - 4 R^{\mu\nu} R_{\mu\nu} + R^{\mu\nu\rho\sigma} R_{\mu\nu\rho\sigma}$ is the four-dimensional Euler density, and the last term in Eq.~\eqref{eq:anomaly_general} is the Pontryagin density, the only possible parity-odd contribution to the trace anomaly. Since any local relativistic quantum field theory obeys CPT invariance, its coefficient must be purely imaginary, which then would violate unitarity of the theory \cite{Bonora:2014qla} for general background geometries. In this sense this anomaly could be as harmful as t'Hooft anomalies which have also recently been discussed in connection with the effective field theory of the standard model \cite{Cata:2020crs,Bonnefoy:2020tyv,Feruglio:2020kfq,Passarino:2021uxa}.

One of the major difficulties in the study of the conformal anomaly for a Weyl fermion is the implementation of the chiral matrix $\gamma_*$ (in four dimensions usually called $\gamma_5$) together with dimensional regularization \cite{Bollini:1972ui,tHooft:1972tcz, Bollini:1973wu}, since not all the properties of $\gamma_*$ can be generalized from four to arbitrary dimensions \cite{Baikov:1991qz}. Within a consistent prescription for the treatment of $\gamma_*$, one can either preserve cyclicity of the $\gamma$ matrix trace and the relation $\gamma_*^2 = \1$, but at the price of introducing a nonvanishing anticommutator of $\gamma_*$ with the other $\gamma_\mu$'s (with a simple anticommutator that breaks Lorentz invariance \cite{tHooft:1972tcz, Breitenlohner:1977hr} or a complicated one that mantains it \cite{Thompson:1985uv}), or alternatively break the cyclicity of the trace while preserving the anticommutator $\{ \gamma_*, \gamma_\mu \} = 0$ \cite{Korner:1991sx,Kreimer:1989ke}.

The recent work that started the debate is the result of Bonora et al. \cite{Bonora:2014qla}, who computed the trace anomaly using Feynman diagrams, computing directly the expectation value of the stress tensor in a perturbative expansion around Minkowski space. They employed dimensional regularization and a $\gamma_*$ that --- as far as we can see --- does not fall into one of the abovementioned categories. Focusing on the subset of one-loop triangle diagrams, they obtained
\begin{equation}
f_{\text{Bonora et al.}} = \frac{\mathi}{96 (4\pi)^2} \eqend{.}
\end{equation}
This result has been reproduced in \cite{Bonora:2017gzz} by Bonora et al., introducing a new technique called metric-axial-tensor background, in which the spacetime metric is allowed to be composed of two terms: one proportional to the identity and the other to $\gamma_*$. This is a generalization of the approach developed by Bardeen to investigate the coupling of Dirac fermions to vector and axial gauge fields \cite{Bardeen:1969md}. Additionally, some points of the original derivation are clarified both in that article and in Ref. \cite{Bonora:2015nqa}.

On the other hand, the group of Bastianelli et al. has found a vanishing $f$ coefficient. In a first work, they used Pauli--Villars regularization, after which the one-loop computation can be reduced to a Fujikawa-like calculation \cite{Bastianelli:2016nuf}. Afterwards, they have extended their Pauli--Villars regularization to consider a metric-axial-tensor background (i.e., the idea developed by Bonora et al. \cite{Bonora:2017gzz}), obtaining once more a vanishing contribution for the parity-odd term \cite{Bastianelli:2019zrq}. Their result also includes the fact that the (parity-even) conformal anomaly in the Weyl case is half the one for a Dirac fermion.

A vanishing parity-odd anomaly has also been obtained in \cite{Frob:2019dgf} by one of the authors of the present manuscript; in that case, a Hadamard subtraction has been introduced for Weyl fermions. These computations confirm in a completely independent approach the results by Bastianelli et al.

Lately, both the results of Bastianelli et al. and Fr{\"o}b et al. have been criticized by the group of Bonora \cite{Bonora:2019dyv}, alleging that the employed methods do not reproduce the well-studied chiral anomalies of Weyl fermions, and would therefore be inappropriate to analyze the conformal anomaly. In order to pursue further the discussion, we have decided to redo the direct computation using Feynman diagrams \cite{Abdallah:2021eii}, paying attention to possible pitfalls.

\section{The trace anomaly computation using the Breitenlohner--Maison scheme for \texorpdfstring{$\gamma_*$}{gamma*}}

We introduce left-handed Weyl fermions as in \cite{Abdallah:2021eii}: they are four-component fermions $\psi$ which are eigenspinors of the chiral projector $\mathcal{P}_+$ (such that their Dirac conjugate $\bar\psi \coloneq \psi \gamma^0$ is an eigenspinor of $\mathcal{P}_-$); in formulae we have
\begin{equation}
\label{eq:weyl_fermion_def}
\psi = \mathcal{P}_+ \psi \coloneq \frac{1}{2} \left( \1 + \gamma_* \right) \psi, \qquad \bar{\psi} = \bar{\psi} \mathcal{P}_- \coloneq \frac{1}{2} \bar{\psi} \left( \1 - \gamma_* \right) \eqend{.}
\end{equation}
Since we work in dimensional regularization, we have to take the action for a free massless Weyl fermion in curved space right from the beginning in $n$ dimensions, which reads
\begin{equation}
\label{eq:action}
S = - \int \bar{\psi} \mathcal{P}_- \gamma^\mu \nabla_\mu \left( \mathcal{P}_+ \psi \right) \sqrt{-g} \total^n x \eqend{.}
\end{equation}
As usual, $g_{\mu\nu}$ denotes the metric of the background curved spacetime, $g$ its determinant and $\nabla_\mu$ the spinorial covariant derivative induced from the Levi--Civita one. From the action \eqref{eq:action}, one obtains the symmetric stress tensor \cite{Forger:2003ut}
\begin{equation}
\label{eq:tmunu}
T^{\mu\nu} = \frac{1}{2} \bar{\psi} \gamma^{(\mu} \nabla^{\nu)} \psi - \frac{1}{2} \gamma^{(\mu} \nabla^{\nu)} \bar{\psi} \psi + \frac{1}{2} g^{\mu\nu} \left( \gamma^{\mu} \nabla_{\mu} \bar{\psi} \mathcal{P}_+ \psi - \bar{\psi} \mathcal{P}_- \gamma^{\mu} \nabla_{\mu} \psi \right) \eqend{.}
\end{equation}
Classically, it is both conserved and traceless when the fermion is on-shell, i.e., satisfies the Weyl equation $\gamma^\mu \nabla_\mu \left( \mathcal{P}_+ \psi \right) = 0$ that follows from the action \eqref{eq:action}.

After the quantization of the theory, the stress tensor in \eqref{eq:tmunu} becomes an operator \cite{Freedman:2011hp}, and its VEV contains divergent contributions that need to be regularized and subtracted by the process of renormalization. Therefore, the question is whether its renormalized VEV is still conserved and traceless, namely by the subtraction an anomaly may arise.

For the chiral $\gamma_*$ matrix in dimensional regularization, we employ the consistent Breitenlohner--Maison (BM) scheme \cite{Breitenlohner:1977hr}. As mentioned in the introduction, this is a key point since not all the properties that $\gamma_*$ enjoys in four dimensions can be extended to an arbitrary number of dimensions. Otherwise, we use the same standard perturbative expansion around Minkowski space as \cite{Bonora:2014qla}, setting
\begin{equation}
g_{\mu\nu} \equiv \eta_{\mu\nu} + \kappa h_{\mu\nu} \eqend{,}
\end{equation}
where $\eta_{\mu\nu}$ is the flat metric and $\kappa = \sqrt{16 \pi G_\text{N}}$ with Newton's constant $G_\text{N}$ is the perturbative parameter. Working to second order in $\kappa$, we thus obtain an expansion for the VEV of the stress tensor of Weyl fermions in the external field $h_{\mu\nu}$, with the terms in the expansion being certain expectation values in the Minkowski vacuum state for the fermions.

\begin{figure}
\begin{center}
\includegraphics[scale=0.25]{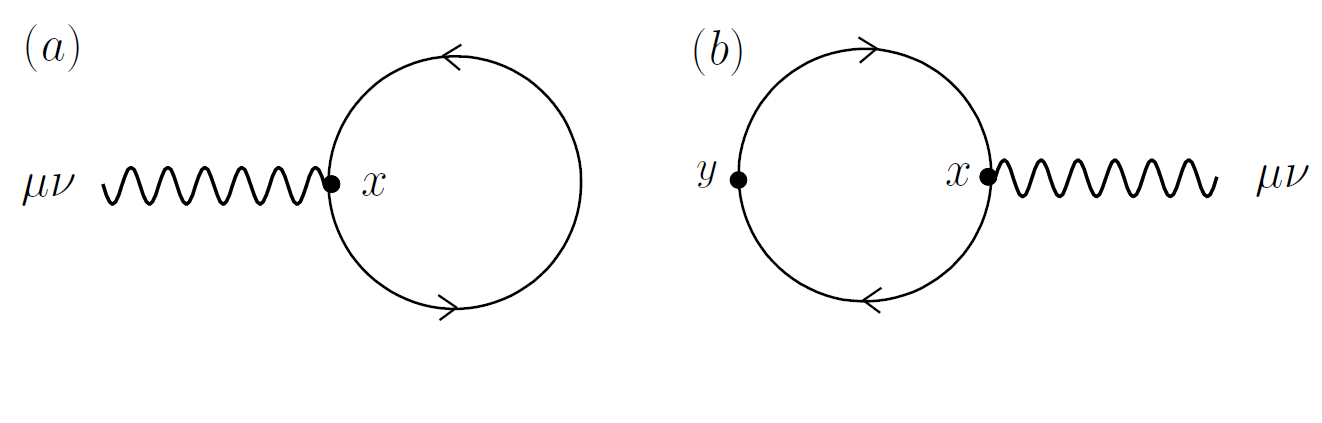}
\caption{Schematics of Feynman diagrams involved in the computation of the anomaly at first order in $\kappa$. Wavy lines correspond to insertions of the external gravitational field, while straight lines are fermion propagators. The VEV of the stress tensor is evaluated at $x$, while $y$ is integrated over.}
\label{fig:feynman1}
\end{center}
\end{figure}

\begin{figure}
\begin{center}
\includegraphics[scale=0.3]{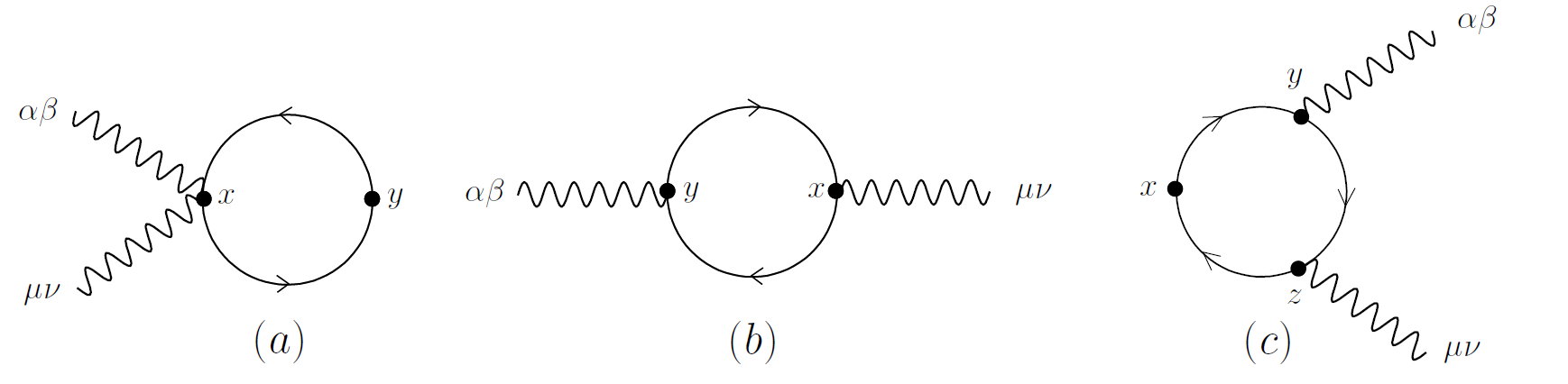}
\caption{Schematics of Feynman diagrams involved in the computation of the anomaly at second order in $\kappa$. Wavy lines correspond to insertions of the external gravitational field, while straight lines are fermion propagators. The VEV of the stress tensor is evaluated at $x$, while $y$ and $z$ are integrated over.} 
\label{fig:feynman2}
\end{center}
\end{figure}

Diagrammatically, the computation involves several contributions, which are schematically displayed in Figs. \ref{fig:feynman1} and \ref{fig:feynman2}, corresponding respectively to first and second order in $\kappa$. The wavy lines correspond to insertions of the external field $h_{\alpha\beta}$, while the straight lines are fermion propagators. The dots generically indicate vertices that could involve scalar as well as tensorial factors; these factors have been omitted to simplify the discussion, and dots in a single diagram do not necessarily correspond to the same type of vertex. Obtaining the actual contribution to the stress tensor involves contractions between all the indices, including those that are not shown.


Since the Riemann tensor is of first order in the metric perturbation $h_{\mu\nu}$ when expanded around flat space, the coefficients of the trace anomaly \eqref{eq:anomaly_general} can be fixed by a computation to second order. Because higher-order contributions become quickly complicated, we thus restrict to second order in $\kappa$. The most challenging computation is the one corresponding to the $\kappa^2$ contribution displayed in Fig. \ref{fig:feynman2}(c), which involves around $10^4-10^5$ terms when expanded in terms of free expectation values of the Weyl fermion in the Minkowski vacuum. For the complicated tensor algebra, we have used the computer tensor algebra suite xAct~\cite{xact,xpert}.

In the computation, it is easy to separate parity-even contributions from parity-odd ones: while the first kind involves an even number of chiral $\gamma_*$ matrices, the second kind involves an odd number of them. Since the identity $\gamma_*^2 = \1$ and the cyclicity of the trace are preserved by the BM scheme, (anti-)commuting $\gamma_*$ with the other $\gamma$ matrices we could reduce the parity-even contributions to $\gamma$ matrix traces without any $\gamma_*$, while the parity-odd contributions could be reduced to a single $\gamma_*$. However, since in the BM scheme $\gamma_*$ anticommutes with $\gamma^0,\ldots,\gamma^3$ and commutes with $\gamma^4,\ldots,\gamma^{n-1}$, Lorentz invariance is explicitly broken and one has to introduce the purely four-dimensional metric $\bar\eta_{\mu\nu}$ in additional to the $n$-dimensional one $\eta_{\mu\nu}$.

For the parity-even part of the stress tensor VEV, we have shown that the difference between our result and the naive computation (with $\gamma_*$ anticommuting with \emph{all} $\gamma$ matrices) is a finite counterterm, such that we recover the well-known result for the trace anomaly after renormalization. On the other hand, for the parity-odd part all possible contributions vanish for one of two reasons: either one contracts an antisymmetric tensor with a symmetric one, or one has to antisymmetrize a purely four-dimensional expression over five or more indices. The prototype of such an expression is $\epsilon^{[\mu\nu\rho\sigma} \bar\eta^{\alpha]\beta} = 0$, since in the BM scheme the antisymmetric Levi--Civita tensor is also purely four-dimensional. Such expressions (and the resulting relations) are known in general as dimensionally dependent identities \cite{Edgar:2001vv}. Since the full stress tensor VEV does not have a parity-odd part, there is also no parity-odd contributions to its trace anomaly. For further details, we refer to \cite{Abdallah:2021eii}.

\section{Conclusions and outlook}

We have computed the full renormalized stress tensor VEV for Weyl fermions, up to second order in an expansion around Minkowski spacetime. The renormalized VEV is conserved, and its trace anomaly is of the form \eqref{eq:anomaly_general} with the coefficients
\begin{equation}
w = - \frac{18}{720 (4 \pi)^2} \eqend{,} \quad b = - \frac{12}{720 (4 \pi)^2} \eqend{,} \quad c = \frac{11}{720 (4 \pi)^2} \eqend{,} \quad f = 0 \eqend{.}
\end{equation}
Since the coefficient of the parity-odd Pontryagin density vanishes, no unitarity problem arises in the quantum theory. The result for the trace anomaly is half of the result for a Dirac fermion, in agreement with \cite{Bastianelli:2016nuf,Bastianelli:2019zrq,Frob:2019dgf}, but contrary to \cite{Bonora:2014qla,Bonora:2019dyv}. It thus seems that the prescription for $\gamma_*$ that was employed in the latter works is not consistent.

It is of course possible to perform further verifications of this result. On one hand, one may try to implement the more general prescription for $\gamma_*$ introduced in \cite{Thompson:1985uv} to check once more whether the parity-odd term of the trace anomaly arises or not; however, the computations will naturally be much more involved. Alternatively, one could use the Breitenlohner--Maison scheme for $\gamma_*$ and verify that the Kimura-Delbourgo-Salam anomaly~\cite{Delbourgo:1972xb} for the axial current in presence of a background gravitational field is reproduced. Possibilities along these lines are under study.

\subsection*{Note added} 
Short after the publication of this work, some 
preprints appeared with partially conflicting results regarding the chiral trace
anomaly~\cite{Bastianelli:2022hmu, Liu:2022jxz}. On the other
hand, we learned from R. Delbourgo that a computation of the
gravitational contribution to the chiral anomaly 
in dimensional regularization has been
performed in~\cite{Delbourgo:1977kt}.

\begin{acknowledgments}
SAF is grateful to G. Gori and the Institut für Theoretische Physik, Heidelberg, for their kind hospitality.
This work has been funded by the Deutsche Forschungsgemeinschaft (DFG, German Research Foundation) --- project nos. 415803368 and 406116891 within the Research Training Group RTG 2522/1.
SAF acknowledges support by UNLP, under project grant X909 and ``Subsidio a J\'ovenes Investigadores 2019''.
\end{acknowledgments}

\bibliographystyle{JHEP}
\bibliography{biblio}

\end{document}